\documentclass[a4paper,fleqn]{cas-dc}

\newcommand{\Fig}[1]{Fig.~\ref{#1}}
\newcommand{\fig}[1]{\Fig{#1}}
\newcommand{\figp}[1]{(\Fig{#1})}

\newcommand{\app}[1]{Appendix~\ref{#1}}
 
\usepackage[authoryear]{natbib}

\def\nat{\rm Nature}
\def\apj{\rm ApJ}
\def\psj{\rm PSJ}
\def\aj{\rm AJ}
\def\icarus{\rm Icarus}
\def\apjl{\rm ApJL}
\def\mnras{\rm MNRAS}
\def\ssr{\rm SSR}
\def\aap{\rm A\&A}
\def\jrasc{\rm Journ.~Royal~Astron.~Soc.~of~Canada}

\begin{document}
\let\WriteBookmarks\relax
\def\floatpagepagefraction{1}
\def\textpagefraction{.001}

\shorttitle{Kuiper belt mass gap}    
\shortauthors{Lyra}  
\title[mode = title]{Where are the missing Kuiper Belt binaries?}  
\author[1]{Wladimir Lyra}[orcid=0000-0002-3768-7542]
\cormark[1]
\ead{wlyra@nmsu.edu}
\ead[url]{http://astronomy.nmsu.edu/wlyra/}
%\credit{}
\affiliation[1]{organization={New Mexico State University, Department of Astronomy},
            addressline={PO Box 30001 MSC 4500}, 
            city={Las Cruces},
            postcode={88001}, 
            state={NM},
            country={USA}}
\cortext[1]{Corresponding author}

\begin{abstract}
In this letter, we call attention to a gap in binaries
in the Kuiper belt in the mass range between $\approx$10$^{19}$-10$^{20}$ kg, with a corresponding dearth in
binaries between 4th and 5th absolute magnitude $H$. The low-mass end
of the gap is consistent with the truncation of the cold classical
population at 400\,km, as suggested by the OSSOS survey, and predicted
by simulations of planetesimal formation by streaming instability. The
distribution of magnitudes for all KBOs is continuous, which means
that many objects exist in the gap, but the binaries in this range
have either been disrupted, or the companions are too close to the
primary and/or too dim to be detected with the current generation of
observational instruments. At the high-mass side of the gap, the objects
have small satellites at small separations, and we find a trend that as mass
decreases, the ratio of primary radius to secondary semimajor
increases. If this trend continues into the gap, non-Keplerian effects
should make mass determination more challenging. 
\end{abstract}

%% Research highlights
%\begin{highlights}
%\item Mass gap in Kuiper Belt binaries between $10^{19}-10^{20}$\,kg
%\item Low-mass end matches planetesimal formation cutoff; high-mass end unclear.
%\item Gap may reflect disrupted binaries or undetected satellites due to observational limits.
%\end{highlights}

\begin{keywords}
Kuiper belt \sep 
Planetesimals \sep
Binaries \sep 
Origin, solar system \sep 
Planetary formation 
\end{keywords}

\maketitle

\section{Introduction}
\label{sect:introduction}

The Kuiper belt is a region of the solar system populated by small
bodies, remnants from the planet formation process. Following the
discovery of Pluto \citep{Slipher30,Tombaugh46} and Charon
\citep{ChristyHarrington78},
the belt now comprises thousands of objects
\citep[c.f.][]{Barucci+08,Prialnik+20} discovered after 1992 
\citep{JewittLuu93}. With the comprehensive statistics, a picture has
emerged that the current Kuiper belt is a remnant of a much more
massive primordial belt of objects, that has been sculpted by the
evolution of the orbits of the giant planets \citep{Stern96,MorbidelliValsecchi97,SternColwell97,Morbidelli+03,Hahn03,Tsiganis+05,Gomes+05,Morbidelli+08,Batygin+11,Dones+15,Malhotra19,MorbidelliNesvorny20,Bottke+23}. The objects are
classified according to their interaction with Neptune: Plutinos
are objects in the 3:2 resonance with Neptune; the Scattered Disk
comprises objects whose perihelia approaches Neptune (30-35 AU),
having thus wide eccentricity and inclination ranges; the
Classical KBOs have moderately low eccentricity, and orbit between 42 and 48 AU, relatively
undisturbed by Neptune. This class is subdivided into a dynamically hot and a dynamically cold
population \citep{Bernstein+04,Noll+08a,Fraser+10,Petit+11}, with the cold classicals probably forming in situ
\citep{ParkerKavelaars10,Batygin+11} thus representing the only
population of pristine planetesimals in the Solar System. Conversely, the larger KBOs were probably formed inside 25\,AU as
theoretical mechanisms for planetary
  growth (planetesimal and pebble accretion) cease
to yield sufficient mass accretion rates to form them beyond this distance \citep{Johansen+15,Lyra+23,Canas+24}. 

In contrast to the asteroid belt between Mars and Jupiter, Kuiper belt objects show a much higher
proportion of binaries \citep{Noll+08b}, which is expected from their much larger Hill
radii for the same mass. The presence of a companion enables the
determination of dynamical masses (if an orbit
can be fit) and, through stellar occultation
observations \citep{Sicardy+06,Braga-Ribas+13,Brown13a} or, less
accurately, thermal radiometry \citep{Stansberry+08,Muller+10,BrownButler17}, volume measurements can be made;  together, these
allow for determining density. The resulting density {\it vs} radius correlation
\citep{Brown13b,Grundy+15,McKinnon+17} or equivalently density {\it vs} mass
\citep{Noll+20,Canas+24} reveals a trend in which smaller objects tend to be
lower density, whereas the larger objects tend to be higher density,
which is interpreted as a result of porosity removal
\citep{BiersonNimmo19,Noll+20} and/or fundamental differences in composition
\citep{Brown13b,Canas+24}. While much effort has been devoted to explaining
the density trend, a major feature of this plot has received less
attention: a gap over a decade in mass with a dearth of binaries.

In this letter, we are interested in the question: is this mass gap real
(i.e., few binaries in this range exist), or is it a result of observational
bias (i.e., the binaries exist but remain undiscovered). We interpret
it as a missing binaries problem, and advance a few possible
explanations for its existence.

\begin{figure*}
  \centering
  \includegraphics[width=.75\textwidth]{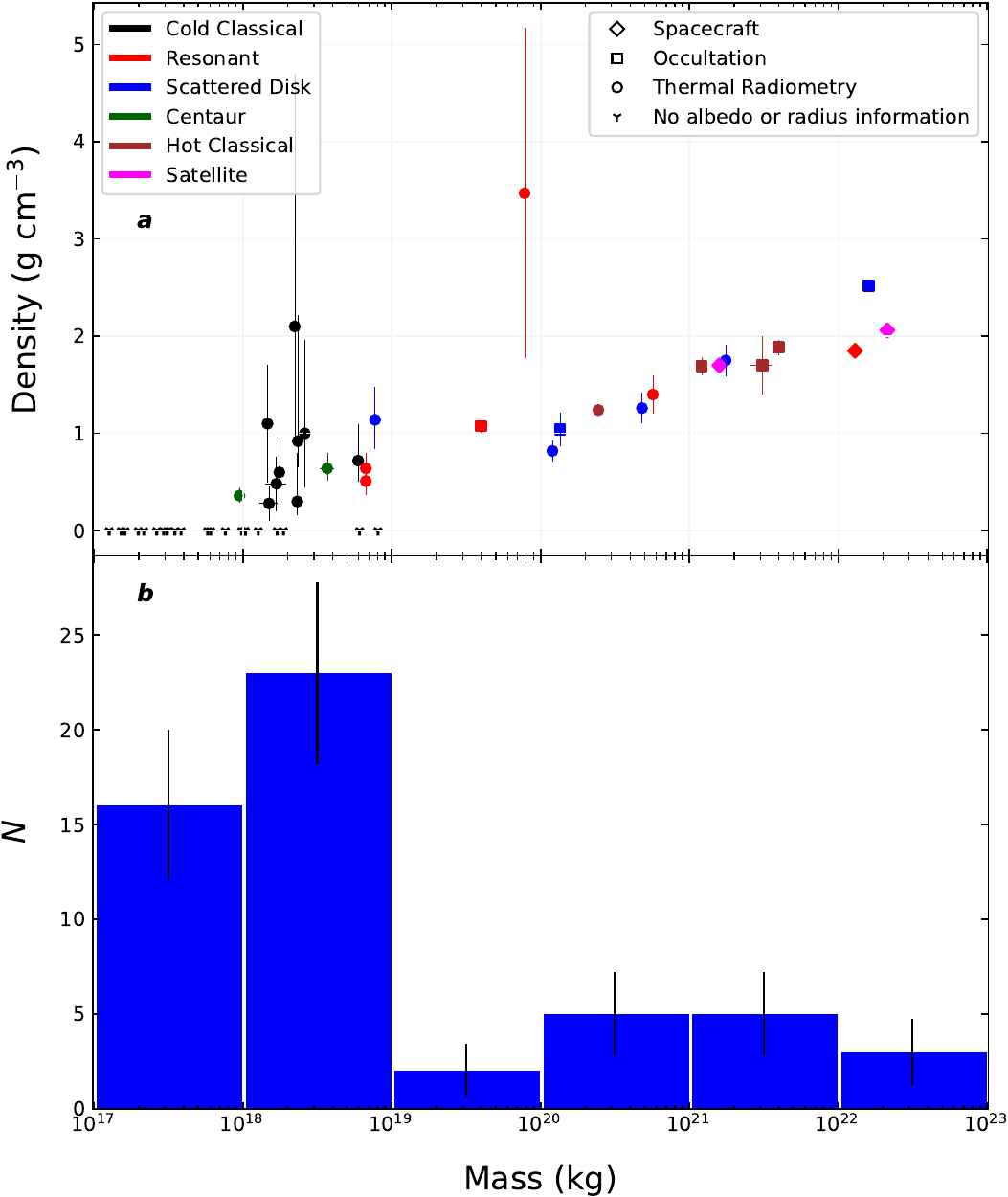}
  \caption{a): The currently known mass vs. density distribution of Kuiper
    Belt objects, showing a dearth of objects in the mass range between
    $\approx 10^{19} - 10^{20}$ kg. Different symbols mark the
      method used for radius determination, and the colors code the
      population the object belongs to. Objects with no radius
      information are plotted at zero density. The two objects in the
      gap are Huya and 2002 WC$_{\rm 19}$; both objects are of higher
      density than the suggested density trend at their masses.
      b): Mass histogram of the same objects as above.}
  \label{fig:kbo_dist}
\end{figure*}

\section{The mass gap}
\label{sect:results}

The gap is illustrated in \fig{fig:kbo_dist}, between roughly $10^{19}-10^{20}$\,kg,
  containing only two objects (2002 WC$_{\rm 19}$ and Huya),
  whereas many objects are present in higher and lower masses. The gap is
also visible in Fig.~3 of \citet{Brown13b}, Fig.~5 of \citet{Grundy+15}, Fig.~1 of \citet{McKinnon+17}, Fig.~1
of \citet{BiersonNimmo19}, and Fig.~1 of \citet{Rommel+25}, in radius, ranging between $\approx$
200-350\,km. Plotting in mass as in the Fig.~5 of \citet{Noll+20} and
Fig.~1 of  \citet{Canas+24} makes the gap more
prominent by cubing the range, to about a decade. The
      data we use is from \citet[][hereafter J19]{Johnston19}, with masses from non-Keplerian orbital fits from
      \citep{Proudfoot+24} when available. We also use updated data for the following
      objects: for Quaoar we use recent occultation data
      \citep{Morgado+23,Pereira+23}; for Gonggong  we use the radius and density data from \citet{Kiss+19}; for Pluto and Charon we use New
  Horizon's updated data from \citet{BrozovicJacobson24};  for Huya we
  use the recent occultation data from \citet{Rommel+25}; for Borasisi
  we use asymmetric error bars for density
$\left(^{+2.6}_{-1.2} \ {\rm g\,cm} ^{-3}\right)$
  from \citet[][hereafter V14]{Vilenius+14}; for 2002 WC$_{\rm 19}$  we used the
  density determination 3.47 $\pm$ 1.7 g\,cm$^{-3}$ from
  \citet{Kovalenko+17} using the thermal radiometry radius from
  \citet{Lellouch+13}. We remove the following
  objects: Mors, because the density of 0.75 g\,cm$^{-3}$ listed in J19 is
  the assumed, not measured, density in \citet{Sheppard+12}; 2001
  QW$_{\rm 322}$, because the reference for density \citep{Petit+08} assumes an
  albedo to find a radius and thus density measurement; 2001 QG$_{\rm 298}$, as
  this system is a contact binary, according to
  \citet{Lacerda11}. The albedo for this object is derived from infrared
  observations, and a shape model is used to derive a density from the
  light curves \citep{Descamps15}. The mass is not
  dynamical, but model-dependent. Finally, we add the following
  objects: 2003 QA$_{\rm 91}$, although the object has no density listed in
  J19, it has dynamical mass, from \citet[][hereafter G19]{Grundy+19}, radius from
  thermal radiometry from V14 and primary-to-secondary radius ratio
  from \citet{Noll+08b}. With this information, we could calculate a
  density of 0.48 ± 0.28 g\,cm$^{-3}$; 2002 VT$_{\rm 130}$, for which we could
  similarly calculate a density of 0.28 ± 0.18 g\,cm$^{-3}$. The masses
  used are the mass of the primary, derived using the system mass
  combined with primary to secondary radius ratio considering equal
  density, a safe assumption for the cold classicals. Equal density is
  not a good assumption for the higher-mass objects, but in that case
  the satellite contributes a small fraction of the system mass, and
  the equal density assumption will not significantly affect the results (we verified that our
  conclusions are unchanged if using system
  mass, see \app{app:systemmass}). For Triton, Pluto, and Charon we use
  their individually determined masses.

\fig{fig:kbo_dist}a shows a density {\it vs} mass plot of the binaries with dynamical mass
determination down to 10$^{17}$ kg; the
  method used for radius determination is shown with different
  symbols, and we color code the different populations. Objects with
  mass but no density information are plotted at zero
  density. \fig{fig:kbo_dist}b is a mass histogram of the same objects
  as \fig{fig:kbo_dist}a. In \fig{fig:kbo_H} we plot the distribution of absolute
magnitudes $H$ of the binaries (blue){\footnote{Obtained from
\url{http://www2.lowell.edu/users/grundy/tnbs/status.html}}}, and the distribution of all KBOs
(red), obtained from the JPL Small-Body
Database{\footnote{\url{https://ssd.jpl.nasa.gov/tools/sbdb_query.html}}}.
A dearth in binaries is seen between magnitudes 4 and
5, whereas the distribution of KBOs magnitudes is continuous. 

\section{Discussion}
\label{sect:discussion}

As noted by \citet{Bernstein+23}, there is a population
difference between the smallest and largest known binary KBOs,
in that the former tend to have more widely separated, equal-sized binaries 
(making them easier to detect), whereas the latter tend to have small satellites on tighter
orbits, suggesting different formation scenarios. Indeed, the
difference is evident in the population breakdown
  show in \fig{fig:kbo_dist}a, in that cold
classicals are all on the low-mass side of the gap.
At the high-mass side of the gap, the objects are primarily
scattered disk objects and hot classicals. The low-mass edge of the gap seems to be consistent
with the cold classical mass distribution: the OSSOS survey \citep{Kavelaars+21}
finds that the distribution of cold classicals is complete to
about 400\,km diameter. At a density of 0.5 g\,cm$^{-3}$, this diameter
corresponds to $\approx 1.68 \times 10^{19}$ kg, consistent
with the beginning of the mass gap. We conclude that the
low-mass end of the gap at $10^{19}$
kg is the high-mass end of the cold classical population, set by the
dynamics of planetesimal formation. 

\begin{figure}
  \centering
  \includegraphics[width=\columnwidth]{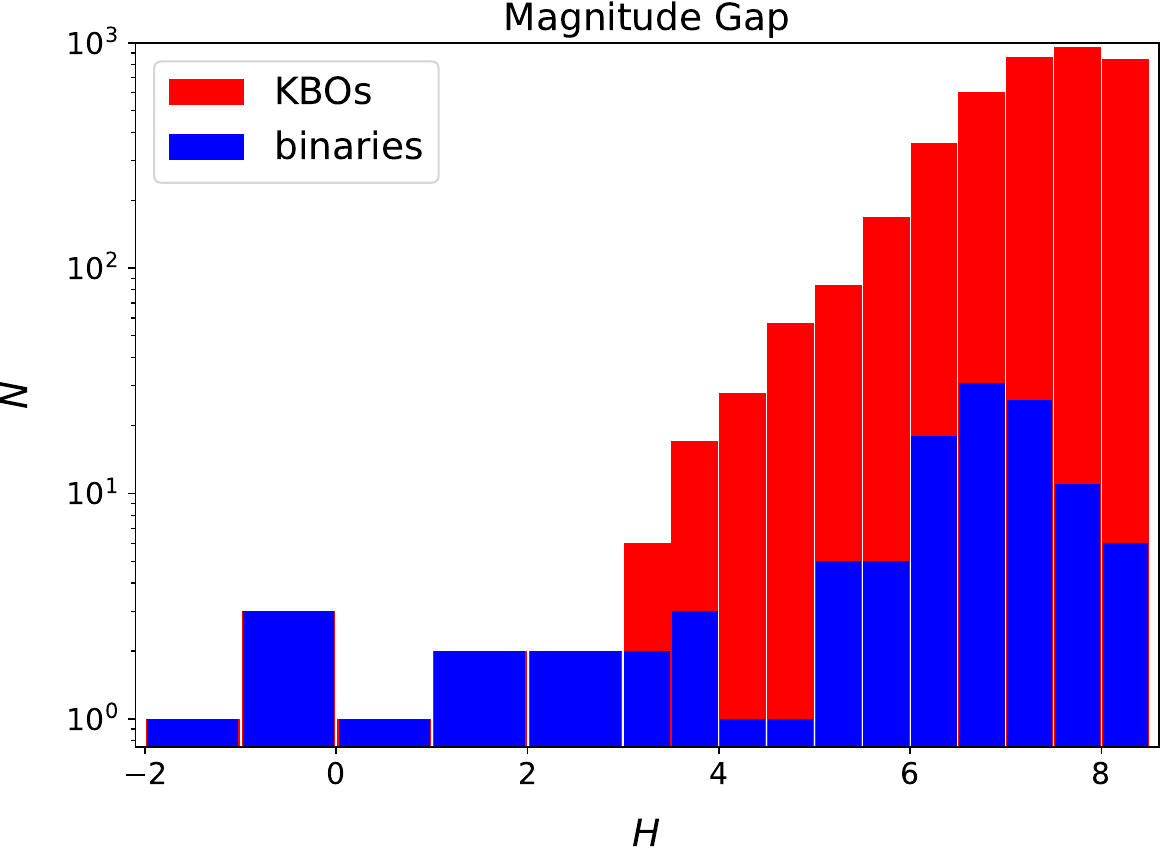}
  \caption{The gap is also present in the distribution of
    magnitudes, as a dearth of binaries between fourth and fifth
    magnitude (blue bins). Note that the distribution of KBOs is
    continuous in this magnitude range, showing that objects in this
    magnitude range (and thus mass range) exist. The two objects in
    the gap are Salacia ($H=4.24\pm 0.02$) and 2000 YW$_{\rm 134}$ ($H=4.72\pm
    0.03$). The two objects in the mass gap (Huya and 2002
      WC$_{\rm 19}$) are on the dim side of the magnitude gap.}
  \label{fig:kbo_H}
\end{figure}

\begin{figure*}
  \centering
  \includegraphics[width=.75\textwidth]{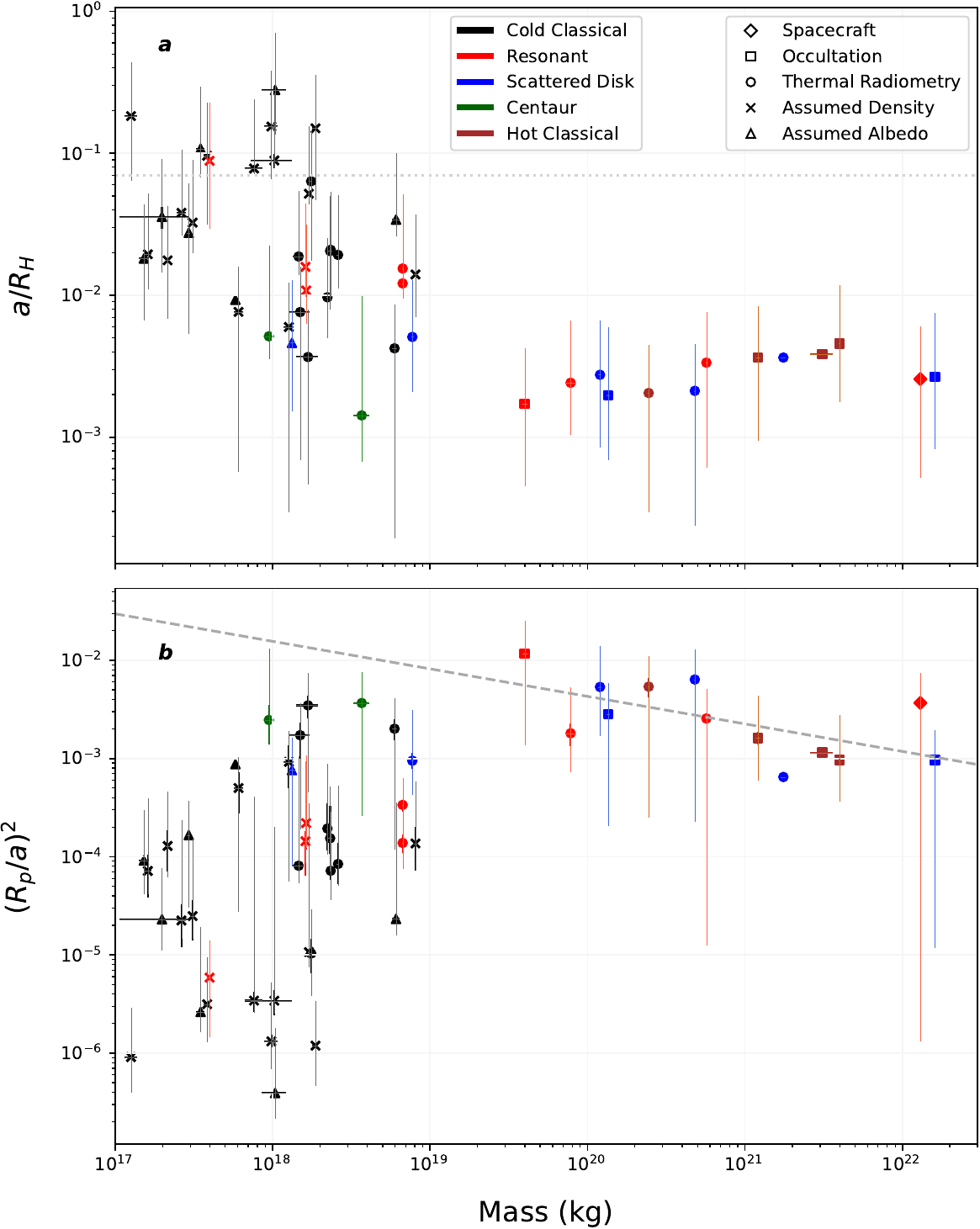}
  \caption{a): Semimajor axis of the secondary as a
    fraction of the binary Hill radius ($a/R_H$). The color code and
    symbols follow the same pattern as \fig{fig:kbo_dist}a, except we split the
    objects with no density information into those for which the
    radius estimate assumed density and those for which the radius
    estimate assumed albedo. Whiskers mark the range of eccentricity
    excursion. The higher mass objects are tight binaries with a flat
    distribution of $a/R_H$, whereas the low mass objects have wider
    separations. The horizontal dotted line at $a/R_H=0.07$ marks the
    transition to ultra-wide binaries. Even though they are cold
    classicals, these ultra-wide binaries are
    not primordial, as they would be dissociated by encounters with
    other planetesimals scattered into the cold classical region
    \citep{Campbell+25}. That the high-mass objects lack
    a similar population of dynamically widened binaries suggests
    that, during their growth, they might have lost their primordial
    satellites. 
    b): Plotting $(R_p/a)^2$ where $R_p$ is the radius of the primary
    and $a$ is the semi-major axis of the secondary, shows decreasing
    trend with mass for the objects in the high-mass side of the mass
    gap, while no trend is seen for the low-mass
    population. The dashed line is the linear regression of
    high-mass objects. If binaries in the gap follow the same trend,
    non-Keplerian effects on the orbit should be
    significant. Huya and 2002 WC$_{\rm 19}$ follow the trend of
    the higher-mass objects in both panels.}
  \label{fig:kbo_J2}
\end{figure*}

The high-mass end of the gap at 
$10^{20}$ kg is less clear. At the high-mass side of
the gap the objects are mostly hot classicals, scattered disk
and resonant bodies. These objects have smaller satellites, of low mass ratio
compared to the primaries. A few possibilities emerge. 

If the binaries are actually missing, we could expect that the binary
disruption was due to the interaction with Neptune, that implanted
the objects in their current orbit in the Kuiper belt. Indeed, these
objects should have formed inside 25\,AU, as pebble accretion is not
efficient beyond it \citep{Lyra+23}. These objects were implanted
by Neptune into their current orbits over about 15\,AU worth of
migration. Given how some cold classicals binaries are weakly bound and
easily disrupted, binaries have to be tightly bound to survive close
encounters with Neptune \citep{ParkerKavelaars10} or with the 
ensemble of planetesimals \citep{Campbell+23,Campbell+25}. 

In \fig{fig:kbo_J2}a we plot the semimajor
  axis of the binary as a fraction of the Hill radius $R_H$. The data
  is taken from J19, and the color and symbol scheme follows that of \fig{fig:kbo_dist}, except that
  now we split the objects with no density information into ``assumed albedo'' and
  ``assumed density''. For the multiples, the semimajor axis of the innermost
  satellite was used: Charon for Pluto, Namaka for Haumea, and Hiisi
  for Lempo\footnote{The Lempo system actually has a 
  third body, Paha, of diameter 132$^{+17}_{-19}$ km. A hierarchical
  triple, their orbits deviate significantly from Keplerian.}.
In addition to the error in $a/R_H$ we add whiskers (thinner
  lines) to mark the eccentricity excursions, both of the orbit of the
  secondary and the binary orbit around the sun. The grey dotted line marks the
  7\% threshold usually quoted for ultra-wide binaries \citep{Parker+11,BruniniZanardi16}. It is
  seen that the high-mass objects are all tight binaries, with $a/R_H$
  between $10^{-3}$ and $10^{-2}$. This is
  at first counter-intuitive in light of the findings of \citet{Campbell+25},
  who show that the ultra-wide binaries started as relatively tighter binaries, and
  got their semi-major axes increased in the first 100 Myr of solar
  system history by encounters with objects from the primordial Kuiper
  belt ejected by the giant planets into crossing the cold classical
  region. Since the disruption depends not on mass but on the $a/R_H$
  ratio \citep{ParkerKavelaars10}, it would be expected that
  low-mass and high-mass objects alike would be similarly affected,
  and a population of wider binaries should also be seen among the higher-mass
  objects.

  A solution to this problem could rest on the pebble accretion
  process. The small objects are
  planetesimals, too low-mass to undergo pebble accretion. In constrast,
  the higher mass objects started as planetesimals at a closer
  distance to the Sun than they are now, and underwent significant
  growth by pebble accretion. Starting from a uniform distribution of
  $a/R_H$ vs mass, pebble accretion might have disrupted by dynamical friction the original binaries
  of the now high-mass objects. This scenario would support the
  present model in which the current satellites of the high-mass Kuiper belt objects are the
  result of collisions
  \citep{Canup05,Brown+06,Canup11,BrownButler18,Arakawa+19,Bernstein+23,BrownButler23}. To
  date, no study of dynamical friction due to pebble accretion on the orbits of
  binaries has been undertaken.

Other effects could also result from non-Keplerian motion. We show in
the lower panel of \fig{fig:kbo_J2} the square of the ratio of primary radius to semi-major
axis of the secondary, $(R_p/a)^2$, which we take as proxy for
non-Keplerian motion, as this is the distance dilution factor for the
quadrupole. The data on $R_p$ was also taken from J19. As with
  the upper plot, in addition to the error bars, whiskers
  mark how the excursion in eccentricity from
  apoastron to  periastron affects $(R_p/a)^2$. The
dashed line shows the linear regression for the high-mass binaries.
A trend is suggested, of larger
quadrupole influence as the mass decreases. Inside the mass gap, the
$(R_p/a)^2$ factor would range from about $5\times 10^{-3}$ to a
little under $10^{-2}$. Non-Keplerian effects do not necessarily lead
to binary disruption, but they do complicate mass determinations,
so it is possible that more binaries in the gap may have been
detected but we simply do not know their masses yet.

The other possibility, of observational bias, is that binaries in the gap exist, with either a
large companion that is nevertheless too close to the primary to
discerne a separation, or with a small satellite that is too dim to
see. The worst case would of course be a satellite that is both too
close to the primary and too dim. Given that the Hill radius is a
slower function of mass than magnitude ($m^{1/3}$ vs $m^{2/3}$,
respectively, where $m$ is mass), a decrease in
mass will thus be more pronounced in magnitude than in
separation. Therefore, the hypothesis that the satellites exist but
are too dim to observe may be the correct one. 

As for the objects in the gap, Huya and 2002 WC$_{\rm 19}$, we
  pose the question: are they
  more similar to the low-mass objects, or to the high-mass objects?
  Given \fig{fig:kbo_dist}a, these objects
  do not follow the density trend of $\approx$0.5 g\,cm$^{-3}$ for low-mass
  objects, and increasing slightly due to gravitational pressure
  \citep{BiersonNimmo19} and/or silicate pebble accretion
  \citep{Canas+24}. Huya has higher density (1.073 $\pm$ 0.066 g\,cm$^{-3}$)
  than 2002 UX$_{\rm 25}$ (0.82 $\pm$ 0.11 g\,cm$^{-3}$), an object about 3 times its
  mass. 2002 WC$_{\rm 19}$  is a more serious outlier, at an unplausible high
  density of 3.5 $\pm$1.7 g\,cm$^{-3}$, similar to fully compact 
  silicates (albeit with a high uncertainty). Determining radius from
  the thermal radiometry technique is not without issues. A lower
  albedo could get the density under 2 g\,cm$^{-3}$, but an unlikely
  low albedo would be needed to bring it to the $\lesssim$ 1 g
  cm$^{-3}$ values of the density trend (the same issue applies to
  Borasisi, which, at $2.1^{+2.6}_{-1.2} \ {\rm g\,cm^{-3}}$, also
  looks to be high density compared to its cold classical cohort). On the
  other hand, as \fig{fig:kbo_J2} shows, Huya and 2002 WC$_{\rm 19}$
  follow very well the trend suggested by the high-mass objects, both
  for the anti-correlation of mass vs $(R_p/a)^2$, and the flat
  $a/R_H$. These characteristics suggest that Huya and 2002 WC$_{\rm 19}$
  could be fragments of differentiated parent bodies or, if
  primordial, formed early enough to undergo melting and porosity
  removal from radiogenic heating.

We note that the gap in magnitude \figp{fig:kbo_H}, although less clear
than the mass gap, is consistent with a bimodal distribution. The objects
inside the $4<H<5$ gap are Salacia ($H=4.24\pm 0.02$), a scattered
disk object on the high-mass side of the mass gap,
and 2000 YW134 ($H=4.72\pm 0.03$), an object in 8:3
resonance with Neptune. Curiously, Salacia is dimmer than 3 objects
with lower mass: 2002 UX$_{\rm 25}$ ($H=3.88\pm 0.01$), Varda ($3.81\pm 0.01$),
and G!k\'un||'h\`omd\'im\`a ($H=3.45\pm 0.03$). As for 2000 YW134,
it does not seem to have a dynamical mass
determination. Yet, the orbital period estimate of
the satellite, 10 days, coupled with a separation of $\approx$1900\,km
\citep{StephensNoll06}, implies a mass around $5\times 10^{18}$
kg, and thus on the low-mass side of the mass gap, and the actual mass is
probably even smaller (W. Grundy, private
communication). We note that Huya ($H=5.31\pm 0.2$) and 2002
  WC$_{\rm 19}$ ($H=5.0\pm 0.5$),  although inside the mass
  gap, are on the dim side of the magnitude gap. This is expected
  given their smaller radii.

  Other interesting cases are Vanth and Dysnomia, the satellites
  of Orcus and Eris, respectively. These are thought to have formed in
  collision events with undifferentiated bodies where the impactor is
  left ``intact'', similar to the event that produced the Pluto-Charon
  system \citep{Canup05,Canup11,Arakawa+19}. Unlike collision events where mass
  from a differentiated body is cast into orbit, forming a
  predominantly icy object of high albedo (like the satellites of
  Haumea), in these events the satellite that remains is of roughly
  similar density as the impactor. Vanth has mass $(8.33\pm
  0.08)\times 10^{19}$ kg \citep{BrownButler23}, putting it in the gap, and density
  $1.5^{+1.0}_{-0.5}$ g\,cm$^{-3}$. Dysnomia has only $1\sigma$ upper
  limits, $1.4\times 10^{20}$ for mass and 1.2 g\,cm$^{-3}$ for
  density \citep{BrownButler23}, which would make it similar to G!k\'un||'h\`omd\'im\`a, but
  more probably lower mass and thus likely in the gap. Vanth is in the magnitude gap, at $H=4.88$; Dysnomia is
  dimmer, at $H=5.6$. Yet, if the collision hypothesis is correct, their masses are not
  primordial, as simulations of the collision events show that the
  resulting satellite is about 3 times less massive than the impactor
  \citep{Canup05,Canup11}. Versions of \fig{fig:kbo_dist} with Vanth and
  Dysnomia at their current and triple mass are shown in \app{app:satellites}. Whatever their primordial mass, these
  objects support the hypothesis that the
  gap is an observational bias. Since they are secondaries, not
  primaries, their detection and mass determination are relatively simpler tasks
  than if they were the primaries accompanied by much smaller
  satellites.

\section{Conclusion}
\label{sect:conclusion}

In this letter, we underscore a gap in binaries
in the Kuiper belt in the mass range between $\approx$10$^{19}$-10$^{20}$ kg, with a corresponding dearth in
binaries between 4th and 5th absolute magnitude $H$. The gap has
appeared in graphs in the literature before, but focus has been put on
density trends or size ratio and separation of the binaries. Many objects
exist in the gap, evidenced by the fact that the $H$ magnitude
distribution is continuous. The origin of the gap is unclear, but
evidence points to it being a combination of formation imprint and
observational bias. The low-mass end of the gap is consistent
with the truncation of cold classicals at 400\,km diameter, thought to
be primordial, resulting from formation by streaming instability, which does show
an exponential tapering at the high-mass end of the initial mass
function. Cold classicals binaries are usually of equal mass, making
them easy to detect. As such, we do not expect future discoveries of
binaries in the gap to be equal brightness as common among the cold
classicals; indeed, the objects in the gap, Huya and 2002
  WC$_{\rm 19}$, follow the trend of the higher-mass objects of
  small satellites in tight orbits. The system of Huya and its
satellite is similar to Orcus-Vanth in size ratio, and
2002 WC$_{\rm 19}$ is about 4 times larger than
its satellite. They are also tight binaries, with $a/R_H$ 0.2\% and 0.4\%, respectively.\

At the high-mass side of the gap, the objects
have small satellites, which poses an observational challenge as at
low mass these small satellites would have to be too close to the
primary to have survived interaction with either Neptune
or the primordial belt, and possibly also
too dim to detect with current observational
capabilities. Mass determinations of objects in the gap may also be
plagued by non-Keplerian effects given the trend of increasing
$(R_p/a)^2$ with decreasing mass. We end this study then with the
question: do smaller non-cold classicals objects have satellites? In a future study we will constrain the observational
parameter space, in terms of magnitude and separation, that would be
needed to discover these missing binaries, if they exist. 

\section*{Acknowledgements}

I acknowledge discussions with Mike Brown, Casey Lisse, Marc Buie,
Daniel Carrera, and Will Grundy. The mass gap was first pointed out to me by Kaitlin
Kratter, as a question following a talk where I showed Fig.~1 of
\citet{Canas+24}. I thank the two anonymous referees for
  comments that helped improve the manuscript. This research was made possible by the
NASA Emerging Worlds program, via grant No. 80NSSC22K1419.

\appendix
\section*{Appendix}

\section{System mass vs primary mass}
\label{app:systemmass}

\begin{figure*}
  \centering
  \includegraphics[width=.495\textwidth]{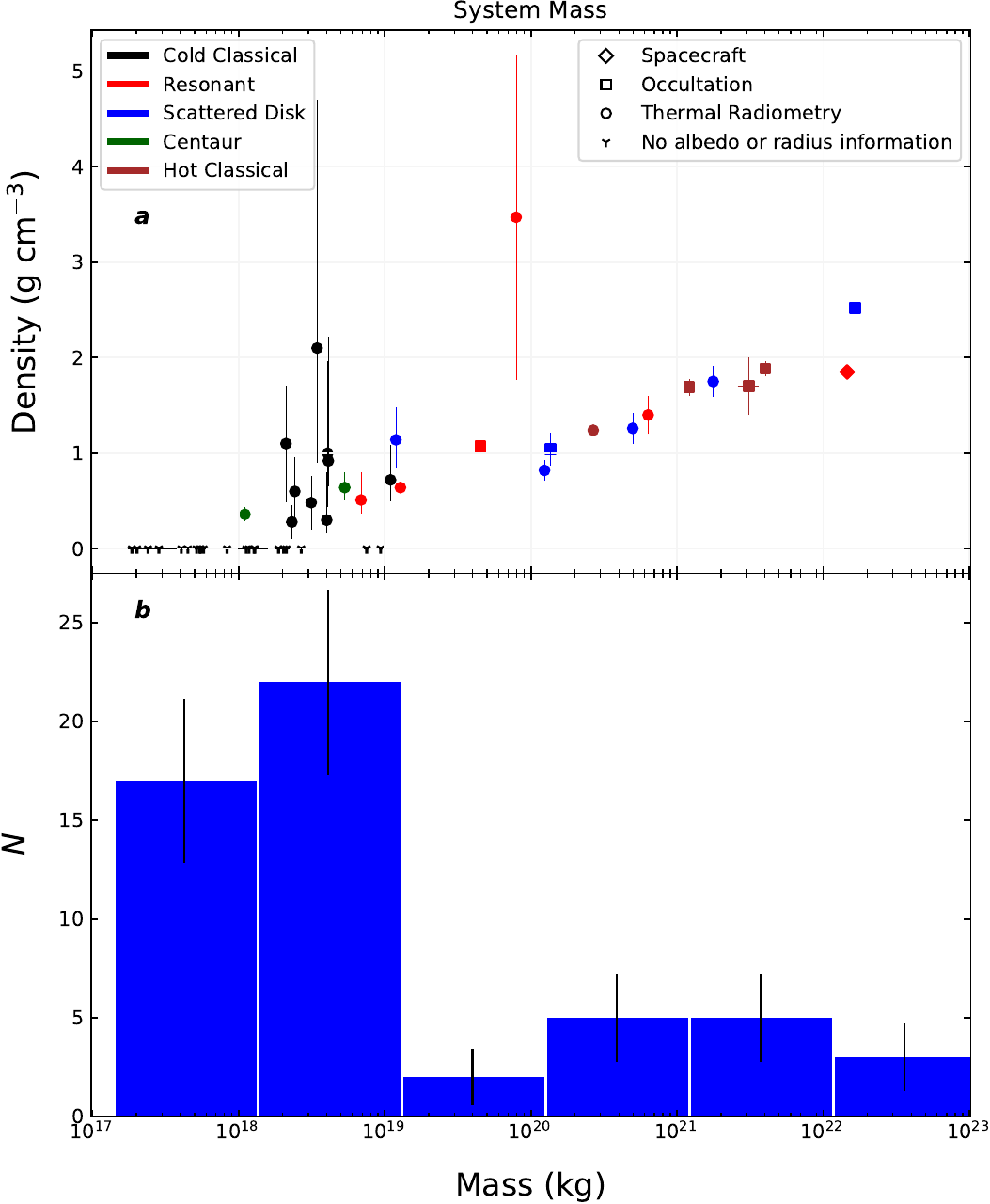}
  \includegraphics[width=.495\textwidth]{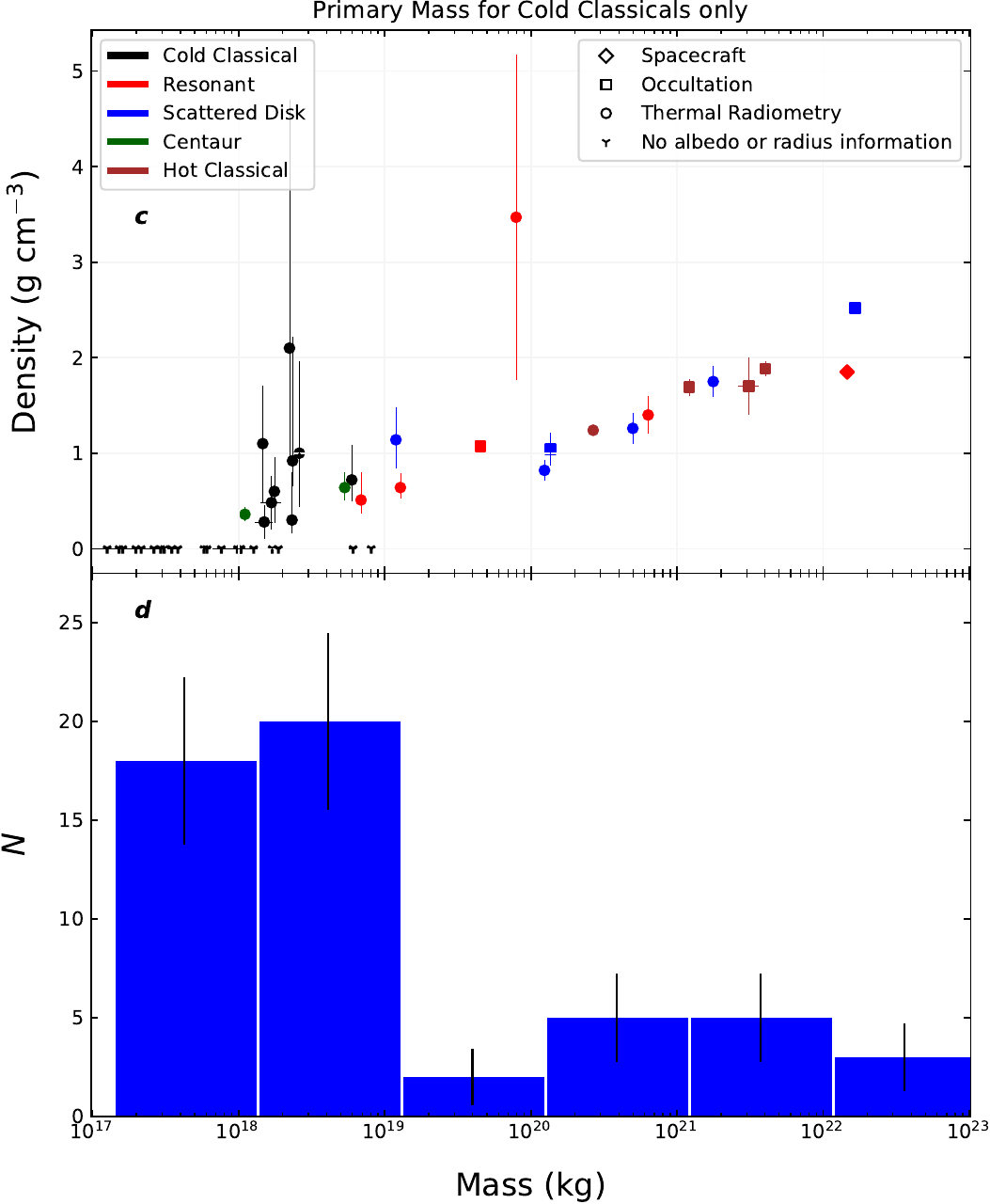}  
  \caption{a and b) Same as Fig 1a and Fig 1b but for system
    mass. The bin size is the mass interval between Lempo and 2002
    UX$_{\rm 25}$. c and d) Same as a and b but for primary mass only
    for the cold classicals. Satellites are excluded.}
  \label{fig:kbo_sysmass}
\end{figure*}

Given the ambiguity on using system mass, or primary mass (which needs
information on radius ratio and assumption of equal density), we show
here as supplementary information a version of \fig{fig:kbo_dist} with system
mass (\fig{fig:kbo_sysmass}, panels a and b). We also show an analysis where primary mass was used only for
the cold classicals and system mass for the other objects (\fig{fig:kbo_sysmass}, panels c and d), motivated
by the fact that equal density is a safe assumption for the cold
classicals only, as their equal colors evidence a common origin. For these cases, satellites (Triton
and Charon) are removed, and Charon's mass is added to Pluto's.
Conclusions are unchanged provided that, when using system mass, we
change the bin interval and range from decimals numbers
(10$^{19}$-10$^{20}$ kg) to between the masses of Lempo and 2002
UX$_{\rm 25}$.

\section{Satellites}
\label{app:satellites}

\begin{figure*}
  \centering
  \includegraphics[width=.495\textwidth]{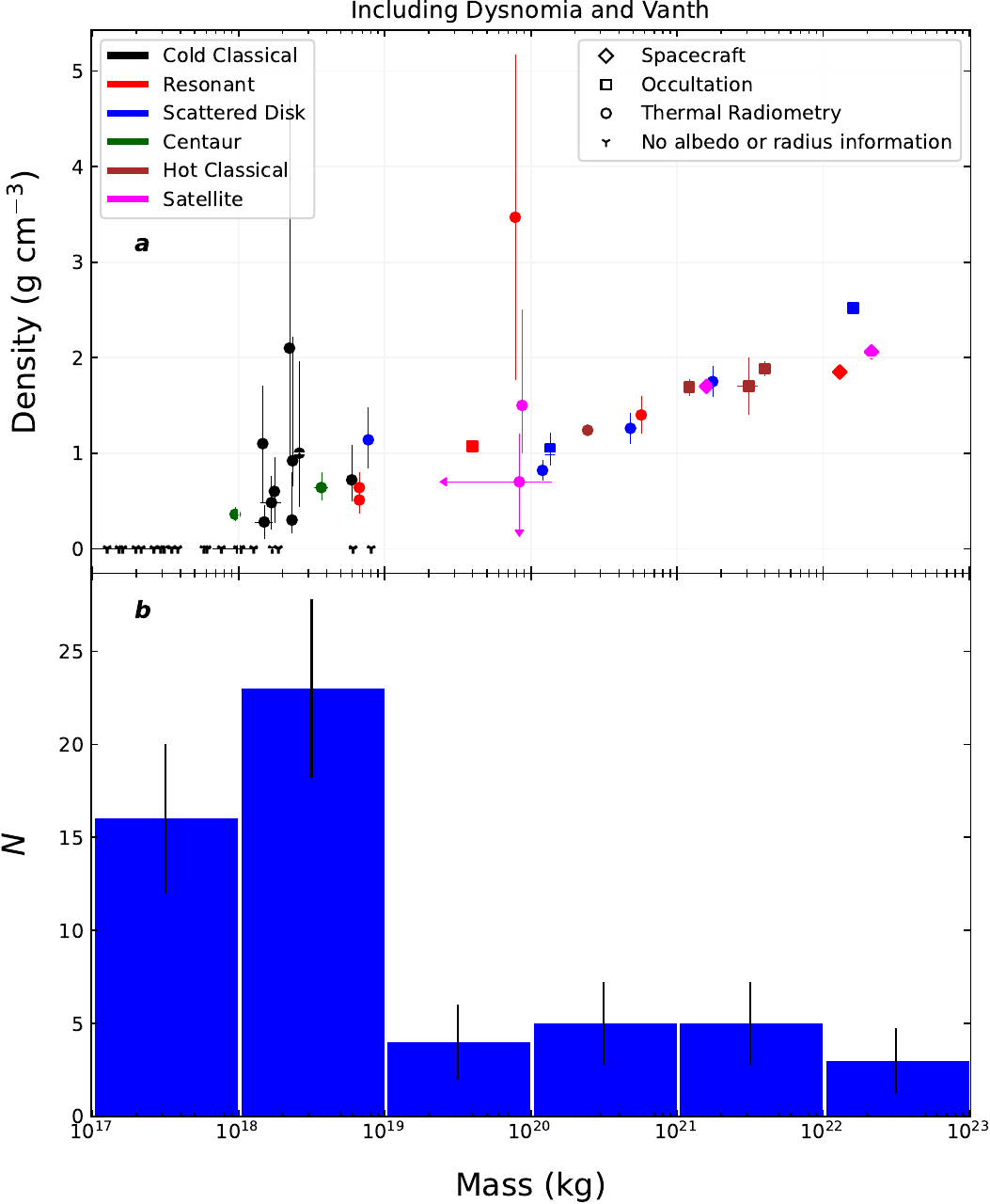}
  \includegraphics[width=.495\textwidth]{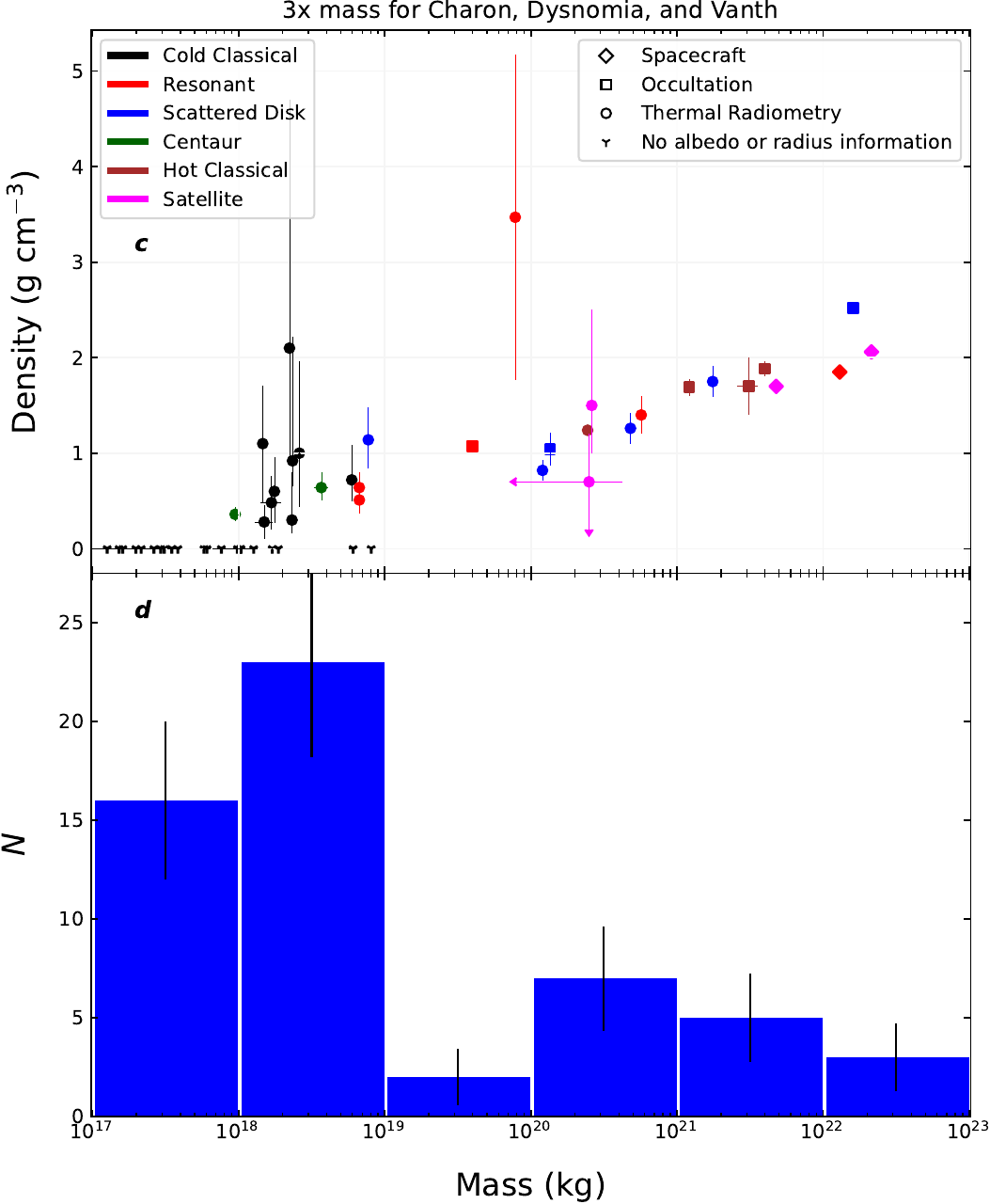}
  \caption{a and b) Same as Fig 1a and Fig 1b but including Vanth and
    Dysnomia. c and d) Same as a and b but tripling the mass of
    Charon, Vanth, and Dysnomia.}
  \label{fig:kbo_satellites}
\end{figure*}

Given recent mass and radius determination of Vanth and an upper limit for
  Dysnomia \citep{BrownButler23} we show in \fig{fig:kbo_satellites},
  panels a and b, a  version of \fig{fig:kbo_dist} including them. Their masses fall in
  the gap; Vanth does not follow the density trend (having similar
  density to Orcus), whereas Dysnomia does, within the upper part of
  the uncertainty range. \fig{fig:kbo_satellites} panels c and d,
  are the same as a and b but for three times the mass of Charon,
  Dysnomia, and Vanth, considering the typical outcomes of the mass
  ratio of impactor to resulting satellite in the ``intact fragment''
  collision model \citep{Canup05,Canup11}. Triton retains its mass
  since it was likely captured via 3-body interaction
  \citep{AgnorHamilton06}, instead of a giant impact. With these
  modified masses, proto-Charon continues following
  the density trend, proto-Vanth now follows it, and proto-Dysnomia does
  not. If they are indeed intact fragments from an undifferentiated
  impactor, their densities being similar to the impactor, then this
  would imply that Vanth lost significant mass, whereas Dysnomia lost
  little. Such conclusion is highly uncertain, though, since Dysnomia
  seems to be in a regime intermediate between intact fragment and
  disk reaccretion \citep{Arakawa+19}.

%\bibliography{master}{}
%\bibliographystyle{cas-model2-names}

\end{document}